\documentclass[useAMS,usenatbib]{mn2e}
\usepackage{txfonts}
\usepackage{natbib}
\usepackage[all]{xy}
\usepackage[dvips]{graphicx}

\def\del#1{{}}

\newcommand{\ltsima}{$\; \buildrel < \over \sim \;$}
\newcommand{\lsim}{\lower.5ex\hbox{\ltsima}}
\newcommand{\gtsima}{$\; \buildrel > \over \sim \;$}
\newcommand{\gsim}{\lower.5ex\hbox{\gtsima}}
\newcommand{\bra}{\langle}
\newcommand{\ket}{\rangle}

\newcommand{\dd}{\mathrm{d}}

\newcommand{\vecl}{\bmath{L}}

\newcommand{\vecx}{\bmath{x}}
\newcommand{\vecr}{\bmath{r}}
\newcommand{\veck}{\bmath{k}}
\newcommand{\vecq}{\bmath{q}}
\newcommand{\vecv}{\bupsilon}
\newcommand{\vecw}{\bmath{w}}
\newcommand{\vecy}{\bmath{y}}
\newcommand{\lprime}{\ell^\prime}

\newcommand{\mprime}{m^{\prime}}

\newcommand{\nprime}{n^\prime}
\newcommand{\ci}{\mathrm{i}}

\newcommand{\dirac}{\delta_\mathrm{D}}
\newcommand{\trace}{\mathrm{tr}}
\newcommand{\determinant}{\mathrm{det}}

\newcommand{\matrixw}{\mathbfss{W}}
\newcommand{\matrixy}{\mathbfss{Y}}

\newcommand{\matrixr}{\mathbfss{R}}
\newcommand{\matrixs}{\mathbfss{S}}

\newcommand{\matrixx}{\mathbfss{X}}
\newcommand{\matrixi}{\mathbfss{I}}

\newcommand{\amatrix}{\mathbfss{A}}
\newcommand{\bmatrix}{\mathbfss{B}}

\title[Galactic angular momentum couplings]
{Galactic angular momenta and angular momentum couplings in the large-scale structure}
\author[B.M. Sch\"afer and Ph.M. Merkel]
{Bj{\"o}rn Malte Sch\"afer$^{1,2,3}$\thanks{e-mail: spirou@ita.uni-heidelberg.de} and
Philipp M. Merkel$^4$\\
$^1$Institute of Cosmology and Gravitation, University of Portsmouth, Dennis Sciama building, Burnaby Road, Portsmouth PO1 3FX, United Kingdom\\
$^2$Institut d'Astrophysique Spatiale, Universit{\'e} de Paris XI, b{\^a}timent 120-121, Centre universitaire d'Orsay, 91400 Orsay CEDEX, France\\
$^3$Astronomisches Recheninstitut, Zentrum f{\"u}r Astronomie, Universit{\"a}t Heidelberg, M{\"o}nchhofstra{\ss}e 12, 69120 Heidelberg, Germany\\
$^4$Institut f{\"u}r Theoretische Astrophysik, Zentrum f{\"u}r Astronomie, Universit{\"a}t Heidelberg, Albert-Ueberle-Stra{\ss}e 2, 69120 Heidelberg, Germany}

\begin{document}
\pagerange{\pageref{firstpage}--\pageref{lastpage}}
\pubyear{2008}
\maketitle
\label{firstpage}

\begin{abstract}
In this paper, we revisit the aquisition of angular momentum of galaxies by tidal shearing and compute the angular momentum variance $\sigma_L^2$ as well as the angular momentum correlation function $C_L(r)$ from a peak-restricted Gaussian random process. This stochastic process describing the initial conditions treats both the tidal shear as well as the inertia as dynamical fields and explicitly accounts for the discreteness of the inertia field. We describe the way in which the correlations in angular momentum result from an interplay of long-ranged correlations in the tidal shear, and short ranged correlations in the inertia field and which reflects the correlation between the eigensystems of these two symmetric tensors. We propose a new form of the angular momentum correlation function which is able to distinguish between parallel and antiparallel alignment of angular momentum vectors, and comment on implications of intrinsic alignments for weak lensing measurements. We confirm the scaling $L/M\propto M^{2/3}$ and find the angular momentum distribution of Milky Way-sized haloes to be correlated on scales of $\sim 1~\mathrm{Mpc}/h$. The correlation function can be well fitted by an empirical relation of the form $C_L(r)\propto\exp(-[r/r_0]^\beta)$.
\end{abstract}


\section{Introduction}
In the current paradigm, haloes acquire angular momentum by tidal shearing from the ambient matter distribution \citep[][for a review]{1988MNRAS.232..339H, 1996MNRAS.282..436C, 1996MNRAS.282..455C, 2000ApJ...532L...5L, 2006ApJ...644L...5L, 2006astro.ph..6477L, 2009IJMPD..18..173S}, which was first proposed by \citet{1949MNRAS.109..365H} and \citet{1955MNRAS.115....2S}. Tidal shearing is well supported by numerical simulations \citep{1984ApJ...286...38W, 2000MNRAS.311..762S, 2001MNRAS.320L...7C, 2007MNRAS.381...41H, 2010MNRAS.405..274H} and leads to alignments of the angular momentum direction with the local tidal shear field.  An important observational consequence of angular momentum alignments in the large-scale structure, are induced intrinsic ellipticty alignments between neighbouring galaxies \citep{2001MNRAS.323..713C, 2002MNRAS.335L..89J}, which can be expected to be a significant source of systematics in weak lensing surveys \citep{2000ApJ...545..561C, 2000MNRAS.319..649H, 2001ApJ...559..552C, 2004PhRvD..70f3526H, 2005astro.ph..6441K, 2008MNRAS.388..991S} and even galaxy surveys as they introduce selection effects due to correlated angles of inclination of the galactic disks \citep{2011MNRAS.410.2730K}. By now, there is reliable observational evidence of tidal-shearing induced ellipticity correlations \citep{2006MNRAS.367..611M, 2007MNRAS.381.1197H, 2010MNRAS.408..897J, 2011arXiv1101.4017B} in particular with SDSS-data, and confirmations of these alignments in numerical simulations \citep{2007ApJ...655L...5A, 2009arXiv0912.1051B, 2010MNRAS.402.2127S}.

In this paper, we revisit the acquisition of angular momentum of cosmological objects in linear theory and recompute the correlation function of angular momenta. We restrict ourselves to linear structure formation, using the Zel'dovich mapping for the description of the tidal shearing mechanism. In this paper, we hope to improve previous works on this topic in these aspects: 
\begin{enumerate}
\item{We employ an improved functional form for the correlation function which is able to distinguish between parallel and antiparallel alignments of angular momenta, and which may assume negative values for antiparallel orientation of the angular momentum vectors. This requires that the angular momentum correlation function can not be a mere quadratic form in the tidal shear and inertia tensor fields, which are the relevant quantities for angular momentum build-up, but needs to be antisymmetric.}
\item{Both the inertia and tidal shear fields will be consistently computed from a correlated Gaussian random process, such that the fields have consistent phase relations. The angular momentum correlation will reflect the different correlation lengths of the inertia and tidal shear fields.}
\item{Treating both fields as dynamical quantities improves on the parameterisation introduced by \citet{2000ApJ...532L...5L} and \citet{2001ApJ...559..552C} for the average misalignment between the eigensystems of both tensors and reflects changes in average misalignment with increasing distance.}
\item{We explicitly take account of the discrete nature of the inertia field as the random process restricted to galaxy formation sites in the large-scale structure has a different weighting of certain inertia-shear combinations compared to that of continuous fields, i.e. the angular momentum distribution is biased.}
\end{enumerate}

The theory is developed in Sect.~\ref{sect_ll_correlation}, where we outline the Gaussian model used for determining the angular momentum correlations in the large-scale structure and where we propose an improved form of the angular momentum correlation function. In addition, we comment on the influence of dark energy cosmologies on the angular momentum acquisition. The results are presented in Sect.~\ref{result_l_correlation}, where we compute the angular momentum correlation function along with the angular momentum variance, and investigate their mass-dependence, followed by a discussion in Sect.~\ref{sect_summary} where we summarise our main results and comment on the consequences of the improved angular momentum model on intrinsic ellipticity correlations. 

Throughout, the cosmological model assumed is a spatially flat $\Lambda$CDM cosmology with Gaussian adiabatic initial perturbations in the cold dark matter distribution. Choices for the relevant parameter values are: $\Omega_m=0.25$ with $\Omega_b=0.04$, $\Omega_\Lambda=0.75$, $H_0=100\,h\,\mbox{km~}\mbox{s}^{-1}\mbox{Mpc}^{-1}$ with $h=0.72$, $n_s=1$ and $\sigma_8=0.9$.

\section{Formalism}\label{sect_ll_correlation}
This section describes the Gaussian model used for deriving the angular momentum correlations in the large-scale structure. Sect.~\ref{derivation_zeldovich_shear} explains how haloes acquire rotation by tidal shearing and relates the angular momentum $\vecl$ to the inertia $I_{\alpha\beta}$ and gravitational shear $\Psi_{\alpha\beta}$ in the Zel'dovich-approximation. In Sect.~\ref{derivation_joint_gaussian}, we outline a model for deriving the correlations of shear and inertia from the fluctuation statistics of the density field, based on a joint multivariate Gaussian probability density. The covariances take a particularly simple form if expressed in spherical coordinates, as explained in Sect.~\ref{derivation_spherical_coords} and we elaborate on the shape of the correlation matrices in Sect.~\ref{derivation_correlation_shape}. The correlation function $\bra L_\alpha(\vecx) L_{\alpha^\prime}(\vecx^\prime)\ket$ of the angular momenta is determined in Sect.~\ref{derivation_ll_correlation} by integrating out the Gaussian probability density restricted to peaks in the density field. We discuss a technical issue, namely the misalignment in the shear and inertia eigensystems in Sects.~\ref{derivation_misalignment} and~\ref{derivation_cancellation}. Due to the high dimensionality of the integration, we employ a numerical Monte-Carlo integration scheme, as explained in Sect.~\ref{derivation_montecarlo}.

\subsection{Acquisition of angular momentum by tidal shearing}\label{derivation_zeldovich_shear}
\citet{1970Ap......6..320D} and \citet{1984ApJ...286...38W} suggested that the angular momentum of galaxies originates from tidal torquing between the protogalactic region and the surrounding matter distribution prior to collapse. Assuming a non-spherical shape of the protogalactic region, the angular momentum grows at first order and linearly in time in Einstein-de~Sitter universes, whereas in spherical regions, the acquisition of angular momentum is only a second order effect due to convective matter streams on the boundary surface, as shown by \citet{1969ApJ...155..393P}. 

Quite generally, the angular momentum $\vecl$ of a rotating mass distribution $\rho(\vecr,t)$ contained in the physical volume $V$ is given by:
\begin{equation}
\vecl(t) = \int_V\dd^3r\:\left(\vecr-\bar{\vecr}\right)\times\vecv(\vecr,t)\rho(\vecr,t),
\end{equation}
where $\vecv(\vecr,t)$ is the (rotational) velocity of the fluid element with density $\rho(\vecr,t) = \bra\rho\ket(1+\delta(\vecr,t))$ at position $\vecr$ around the centre of gravity $\bar{\vecr}$. In perturbation theory, $\delta\ll 1$ and the density field can be approximated by assuming a constant density $\bra\rho\ket = \Omega_m\rho_\mathrm{crit}$ inside the protogalactic region. Following \citet{1984ApJ...286...38W}, \citet{1996MNRAS.282..436C} and \citet{2001ApJ...559..552C}, we describe the growth of perturbations on an expanding background in Lagrangian perturbation theory: The trajectory of dark matter particles in comoving coordinates is given by the Zel'dovich approximation \citep{1970A&A.....5...84Z}:
\begin{equation}
\vecx(\vecq,t) = \vecq - D_+(t)\nabla\Psi(\vecq)
\rightarrow
\dot{\vecx} = -\dot{D}_+\nabla\Psi,
\end{equation}
which relates the initial particle positions $\vecq$ to the positions $\vecx$ at time $t$. The particle velocity $\dot{\vecx}$ follows from the Zel'dovich-relation by differentiation by the time-variable. The growth function $D_+(t)$ describes the homogeneous time evolution of the displacement field $\Psi$ and contains the influence of the particular dark energy model. In the Lagrangian frame, the expression for the angular momentum becomes
\begin{equation}
\vecl 
= \rho_0 a^5 \int_{V_L}\dd^3q\:\left(\vecx-\bar{\vecx}\right)\times \dot{\vecx}
\simeq \rho_0 a^5 \int_{V_L}\dd^3q \left(\vecq - \bar{\vecq}\right)\times \dot{\vecx},
\end{equation}
where the integration volume is defined in comoving coordinates as well. 
Assuming that the gradient $\nabla\Psi(\vecq)$ of the displacement field $\Psi(\vecq)$ does not vary much across the Lagrangian volume $V_L$, a second-order Taylor expansion in the vicinity of the centre of gravity $\bar{\vecq}$ is applicable:
\begin{equation}
\partial_\alpha\Psi(\vecq) \simeq \partial_\alpha\Psi(\bar{\vecq}) + \sum_\beta (\vecq - \bar{\vecq})_\beta\Psi_{\alpha\beta}, 
\end{equation}
The expansion coefficient is the tidal shear $\Psi_{\sigma\gamma}$ at the point $\bar{\vecq}$:
\begin{equation}
\Psi_{\sigma\gamma}(\bar{\vecq}) = \partial_\sigma\partial_\gamma\Psi(\bar{\vecq}),
\end{equation}
because the Zel'dovich displacement field $\Psi$ is related to gravitational potential $\Phi$ and can be computed as the solution to Poisson's equation $\Delta\Psi=\delta$ from the cosmological density field $\delta$. The gradient $\partial_\alpha\Psi(\bar{\vecq})$ of the Zel'dovich potential displaces the protogalactic object, which is neglected in the further derivation, as we only trace differential advection velocities responsible for inducing rotation. Identifying the tensor of second moments of the mass distribution of the protogalactic object as the inertia $I_{\beta\sigma}$,
\begin{equation}
I_{\beta\sigma} = \rho_0 a^3 \int_{V_L}\dd^3q\: (\vecq - \bar{\vecq})_\beta (\vecq -\bar{\vecq})_\sigma
\end{equation}
one obtains the final expression of the angular momentum $L_\alpha$:
\begin{equation}
L_\alpha = a^2 \dot{D}_+ \epsilon_{\alpha\beta\gamma} \sum_\sigma I_{\beta\sigma}\Psi_{\sigma\gamma}.
\end{equation}
It is convenient to rewrite the time dependence of $D_+$ in terms of the scale factor $a$ by $\dd D_+/\dd t = a H(a)\dd D_+/\dd a$, yielding:
\begin{equation}
L_\alpha = a^3 H(a)\frac{\dd D_+}{\dd a}\:
\epsilon_{\alpha\beta\gamma} \sum_\sigma I_{\beta\sigma}\Psi_{\sigma\gamma}.
\end{equation}
The theory of angular momentum acquisition by tidal shearing has been extended to nonlinear stages by using second order perturbation theory \citep{1996MNRAS.282..455C} and to include effects of non-Gaussian initial perturbations \citep{1997MNRAS.292..225C}, but for reasons of analytical computability, we restrict our model of angular momenta to the linear regime of structure formation of a Gaussian random field.

\begin{figure}
\resizebox{\hsize}{!}{\includegraphics{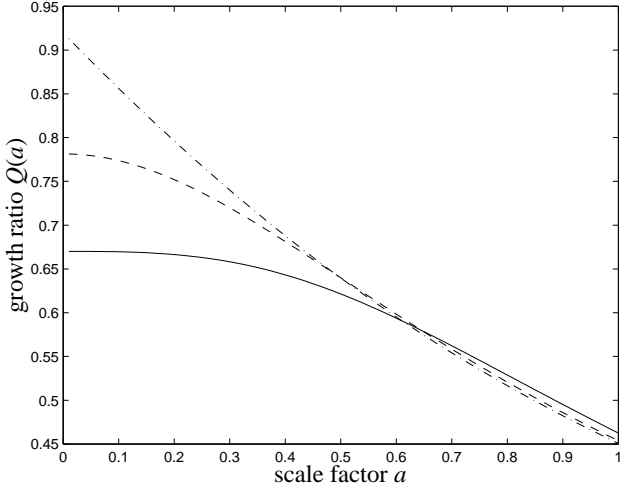}}
\caption{The influence of the dark energy model on the time evolution of angular momenta: $Q(a)$ as a function of scale factor $a$ for $\Lambda$CDM ($w_0=-1$ and $w_a=0$, solid line), for quintessence ($w_0=-2/3$ and $w_a=0$, dashed line) and a dark energy model with variable equation of state ($w_0=-2/3$ and $w_a=1/3$, dash-dotted line).}
\label{fig_time_evolution}
\end{figure}

Fig.~\ref{fig_time_evolution} compares the time evolution of the angular momentum in dark energy cosmologies with SCDM. We define the ratio
\begin{equation}
Q(a)\equiv \frac{q_\mathrm{DE}(a)}{q_\mathrm{SCDM}(a)}
\end{equation}
with $q(a)=a^3 H(a) \dd D_+/\dd a$ (with $D_+(a)$ normalised to unity today and we parameterise the dark energy equation of state with \citep{2001IJMPD..10..213C, 2003MNRAS.346..573L}
\begin{equation}
w(a)=w_0+(1-a)w_a. 
\end{equation}
In SCDM these formulae simplify to $H(a)=H_0a^{-3/2}$, $D_+(a)=a$ and consequently $q_\mathrm{SCDM}=H_0a^{3/2}$. Fig.~\ref{fig_time_evolution} suggests that the spin-up of haloes in dark energy models is significantly slower compared to SCDM, and the choice of the equation of state affects the time evolution significantly. The growth function $D_+(a)$ and its derivative $\dd D_+/\dd a$ follows numerically as a solution to the growth equation,
\begin{equation}
\frac{\dd^2}{\dd a^2}D_+ +\frac{1}{a}\left(3+\frac{\dd\ln H}{\dd\ln a}\right)\frac{\dd}{\dd a}D_+ = \frac{3}{2a^2}\Omega_m(a) D_+(a),
\end{equation}
in which the dark energy model affects the scaling of the Hubble function $H(a)$ and of the matter density parameter $\Omega_m(a)$. In spatially flat dark energy cosmologies, the Hubble function $H(a)=\dd\ln a/\dd t$ is given by
\begin{equation}
\frac{H^2(a)}{H_0^2} = \frac{\Omega_m}{a^{3}} + (1-\Omega_m)\exp\left(3\int_a^1\dd\ln a^\prime\:(1+w(a^\prime))\right),
\end{equation}
with the dark energy equation of state $w(a)$. The value $w\equiv -1$ corresponds to the cosmological constant $\Lambda$.

\subsection{Gaussian model of the angular momentum correlations}\label{derivation_joint_gaussian}
The goal of this section is to derive the 2-point correlation function of the angular momenta of objects that form at peaks in the cosmic  density field. The quantities needed are the the tidal shear $\Psi_{\alpha\beta}(\vecx)$ as well as the inertia $I_{\alpha\beta}(\vecx)$ of a peak region, which both can be related to the density field itself and its second derivatives. In cosmology, fluctuations in the distribution of matter are described by the overdensity $\delta(x)$, which is defined as the fractional perturbation in the density field $\rho(\vecx)$, $\delta(\vecx) = (\rho(\vecx)-\bra\rho\ket)/\bra\rho\ket$, with the average density $\bra\rho\ket=\Omega_m\rho_\mathrm{crit}$. These perturbations are conveniently decomposed in Fourier modes $\delta(\veck)$:
\begin{equation}
\delta(\veck) = \int\dd^3x\: \delta(\vecx)\exp(-\ci\veck\vecx).
\end{equation}
Specifying the power spectrum $P(k)$ suffices to describe the statistical properties of a homogeneous and isotropic Gaussian random field:
\begin{equation}
\bra\delta(\veck)\delta(\veck^\prime)^*\ket = (2\pi)^3\dirac^3(\veck-\veck^\prime) P(k).
\end{equation}
The power spectrum, for which we choose the ansatz $P(k)\propto k^{n_s}\cdot T^2(k)$, is normalised to exhibit a variance of $\sigma_8=0.9$ on scales of $R=8~\mathrm{Mpc}/h$ by the relation:
\begin{equation}
\sigma_R^2 = \frac{1}{2\pi^2}\int\dd k\:k^2 W^2(kR) P(k),
\end{equation}
with a Fourier-transformed spherical top-hat for the filter function $W(y)$, i.e. $W(y)=3\left[\sin(y)-y\cos(y)\right]/y^3$. A common parameterisation for the shape of the transfer function $T(q)$ for CDM models was proposed by \citet{1986ApJ...304...15B}:
\begin{displaymath}
T(q) = \frac{\ln(1+2.34q)}{2.34q}
\nonumber
\end{displaymath}
\begin{equation}
\quad\times\left[1+3.89q+(16.1q)^2+(5.46q)^3+(6.71q)^4\right]^{-\frac{1}{4}},
\end{equation}
where the wave-vector $k$ is given in units of the shape parameter $\Gamma$, first introduced by \citet{1992MNRAS.258P...1E}. A convenient parameterisation of the value of $\Gamma$ as a function of the matter density $\Omega_m$ and the baryonic density $\Omega_b$ is given by \citet{1995ApJS..100..281S}:
\begin{equation}
q = q(k) = \frac{k/\mathrm{Mpc}^{-1}h}{\Gamma}\mbox{ with } \Gamma = 
\Omega_m h\exp\left(-\Omega_b\left[1+\frac{\sqrt{2h}}{\Omega_m}\right]\right).
\end{equation}
The mass scale of the objects of interest is set by imposing a smoothing on high spatial frequencies, where the diameter $R$ of the isotropic filter function $S_R(k)$ corresponds to the size of the objects at the onset of collapse. For numerical reasons, we use a spherically symmetric Gaussian for $S_R(k)$:
\begin{equation}
P(k) \longrightarrow P(k)S_R^2(k)\mbox{ with } S_R(k)=\exp(-k^2R^2/2).
\label{eqn_smoothed_power_spectrum}
\end{equation}
In order to predict the correlation of the angular momenta $\vecl$ of objects which form at peaks in the Gaussian density field, one needs to relate the density gradient $\delta_\alpha(\vecx)$, the second derivatives $\delta_{\alpha\beta}(\vecx)$ and the tidal field $\Psi_{\alpha\beta}(\vecx)$ to the density field $\delta(\vecx)$:
\begin{eqnarray}
\delta(\vecx) & = & 
\int\frac{\dd^3k}{(2\pi)^3}\delta(\veck)\exp(\ci\veck\vecx),\\
\delta_\alpha(\vecx) & = & 
\frac{\partial\delta(\vecx)}{\partial x_\alpha} = 
\ci\int\frac{\dd^3k}{(2\pi)^3}k_\alpha\delta(\veck)\exp(\ci\veck\vecx),\\
\delta_{\alpha\beta}(\vecx) & = & 
\frac{\partial^2\delta(\vecx)}{\partial x_\alpha\partial x_\beta} = 
-\int\frac{\dd^3k}{(2\pi)^3} k_\alpha k_\beta\delta(\veck)\exp(\ci\veck\vecx)\label{eqn_link_inertia_delta},
\end{eqnarray}
The tidal shear follows from the solution of the Poisson equation $\Delta\Psi(\vecx) = \delta(\vecx)$ linking the Zel'dovich potential $\Psi(\vecx)$ to the density field $\delta(\vecx)$:
\begin{equation}
\Psi_{\alpha\beta}(\vecx) =
\frac{\partial^2\Psi(\vecx)}{\partial x_\alpha\partial x_\beta} = 
\int\frac{\dd^3k}{(2\pi)^3} \frac{k_\alpha k_\beta}{k^2} \delta(\veck)\exp(\ci\veck\vecx).
\label{eqn_link_shear_delta}
\end{equation}
An important consequence of eqns.~(\ref{eqn_link_inertia_delta}) and~(\ref{eqn_link_shear_delta}) will be the fact that the angular momentum correlation is determined by two mechanisms with differing correlation length: a short range correlation of the peak shapes and hence the inertia, and a long range correlation mediated by the tidal shear.

The joint distribution of the amplitudes of the density field, its derivatives and the tidal shear follows from a Gaussian probability density \citep{1986ApJ...304...15B}:
\begin{equation}
p(\vecv)\dd\vecv = 
\frac{1}{(2\pi)^{N/2}\sqrt{\determinant\:\mathbfss{V}}}\exp\left(-\frac{1}{2} \vecv^t \mathbfss{V}^{-1}\vecv\right)\dd\vecv,
\label{eqn_derivation_1pt_pdf}
\end{equation}
where the quantities of interest at the point $\vecx$ have been arranged in a 15-dimensional vector $\vecv$, i.e., 3 values for the density gradient $\delta_\alpha(\vecx)$, 6 values for the second derivatives $\delta_{\alpha\beta}(\vecx)$ of the density field (due to the interchangability of the second derivatives) and 6 values for the tidal shear $\Psi_{\alpha\beta}(\vecx)$, which is symmetric under index exchange as well. The covariance matrix $\mathbfss{V}$ follows from the outer product $V_{ij}\equiv\bra\upsilon_i\upsilon_j^*\ket$, $(i,j)=1\ldots15$. This probability density can be extended to include the field values $\vecv^\prime$ at a second point $\vecx^\prime$,
\begin{equation}
p(\vecw)\dd\vecw = 
\frac{1}{(2\pi)^N\sqrt{\determinant\:\matrixw}}\exp\left(-\frac{1}{2} \vecw^t \matrixw^{-1}\vecw\right)\dd\vecw,
\label{eqn_derivation_2pt_pdf}
\end{equation}
where the 30-dimensional vector $\vecw=(\vecv,\vecv^\prime)$ combines the vectors $\vecv$ and $\vecv^\prime$ at the two points $\vecx$ and $\vecx^\prime$ under consideration and the covariance matrix $\matrixw$, $W_{ij}\equiv\bra w_i w_j^*\ket$, $(i,j)=1\ldots30$, is defined in complete analogy.

A peculiarity worthwhile mentioning is the fact that the density field $\delta(\vecx)$ is degenerate with the trace $\trace\Psi_{\alpha\beta}(\vecx)$ of the tidal shear because of Poisson's equation $\Delta\Psi = \trace\Psi_{\alpha\beta}=\delta$. For that reason, the density field will appear in the above outlined random process as a derived quantity, whereas the entries of the tidal shear matrix will be drawn from the Gaussian distribution, with the peak restriction in place.

The inertia tensor $I_{\alpha\beta}$ of an object forming at a peak in the density field at position $\vecx_p$ is related to the second derivatives $\delta_{\alpha\beta}$ of the density field at that particular point \citep{1996MNRAS.282..436C}: In the eigenframe of the mass tensor $-\delta_{\alpha\beta} \equiv- \partial_\alpha\partial_\beta\delta$ at the peak, the density field can be approximated by a parabolic density profile,
\begin{equation}
\delta(\vecx) = \delta(\vecx_p) - \frac{1}{2}\sum_{\alpha=1}^3 \lambda_\alpha (\vecx-\vecx_p)_\alpha^2,
\end{equation}
where $\lambda_\alpha$, $\alpha=1,2,3$ are the eigenvalues of the mass tensor. If the boundary $\partial\Gamma$ of the peak region $\Gamma$ is defined by the isodensity surface $\delta_\Gamma=0$ and if the peak height is expressed in units of the variance $\sigma_0$, $\delta(\vecx)=\nu\sigma_0$, the boundary surface is given in the parabolic approximation by an ellipsoid equation:
\begin{equation}
\partial\Gamma:\:\sum_{\alpha=1}^3\left(\frac{(\vecx-\vecx_p)_\alpha}{A_\alpha}\right)^2 = 1,
\end{equation}
where the semi-axes $A_\alpha$ of the ellipsoid are related to the eigenvalues $\lambda_\alpha$ by
\begin{equation}
A_\alpha = \sqrt{\frac{2\nu\sigma_0}{\lambda_\alpha}}.
\end{equation}
The volume $\Gamma$ of ellipsoidal peak region bounded by isodensity contour $\delta_\Gamma=0$ in the parabolic approximation is then given by:
\begin{equation}
\Gamma = \frac{4\pi}{3}\:A_x A_y A_z,
\end{equation}
which would immediately yield an estimate for the mass of the object:
\begin{equation}
M = \eta_0\Gamma = \eta_0\frac{4\pi}{3}\:A_x A_y A_z.
\label{eqn_mass_parabolic}
\end{equation}
The inertia tensor $I_{\alpha\beta}$ follows from the second moments of the mass distribution, restricted to the volume $\Gamma$ of the peak region, and is diagonal in the mass tensor eigenframe. Carrying out the integration yields:
\begin{equation}
I_{\alpha\beta} 
= \frac{\eta_0}{5}\:\Gamma\:\mathrm{diag}\left(A_y^2+A_z^2, A_x^2+A_z^2, A_x^2+A_y^2\right).
\label{eqn_inertia_parabolic}
\end{equation}
The evolution of the density field is assumed to be homogeneous to first order, $\eta_0\equiv\rho_0a_0^3=\bra\rho\ket a^3$, with $\bra\rho\ket=\Omega_m\rho_\mathrm{crit}$.

\subsection{Describing the correlations in spherical coordinates}\label{derivation_spherical_coords}
Following the example of \citet{1995MNRAS.272..447R} and \citet{1999MNRAS.310.1062H}, we express the correlations between the density field, its derivatives and the tidal shear in spherical coordinates. The two peaks under consideration are assumed to be positioned on the $z$-axis, symmetric about the origin, and separated by a distance $r$, i.e. they have the coordinates $\vecx=(0,0,+z/2)$ and $\vecx^\prime=(0,0,-z/2)$. The correlations take a particularly simple shape in the basis given by the set of dimensionless complex variables $y_{\ell m}^n(\vecx)$:
\begin{equation}
y_{\ell m}^n(\vecx) = \sqrt{4\pi}\frac{\ci^{\ell+2n}}{\sigma_{\ell+2n}}
\int\frac{\dd^3k}{(2\pi)^3}k^{\ell+2n}\delta(\veck)Y_{\ell m}(\hat{k})\exp(\ci\veck\vecx), 
\end{equation}
with $\hat{k}=\veck/k$ as the direction of the wave-vector $\veck$. $\sigma_j^2$ are the weighted moments of the (smoothed) matter spectrum $P(k)$:
\begin{equation}
\sigma_j^2 = \frac{1}{2\pi^2}\int\dd k\:k^{2j+2} P(k).
\label{eqn_derivation_sigmaj}
\end{equation}
The transformation between the physical frame and the $y_{\ell m}^n$-frame for the scalar density field $\delta(\vecx)$ is given by
\begin{equation}
\sigma_0 y_{00}^0(\vecx) = \delta(\vecx).
\end{equation}
For the vectorial density gradient $\delta_\alpha(\vecx�$, they read:
\begin{eqnarray}
\sigma_1 y_{10}^0(\vecx) & = & \sqrt{3}\:\delta_z(\vecx),\\
\sigma_1 y_{11}^0(\vecx) & = & -\sqrt{3/2}\:\left(\delta_x(\vecx)+\ci\delta_y(\vecx)\right).
\end{eqnarray}
The tensor $\delta_{\alpha\beta}(\vecx)$ can be determined from the $y_{\ell m}^n$-coefficients by:
\begin{eqnarray}
\sigma_2 y_{20}^0(\vecx) & = & -\sqrt{5/4}\:\left(\delta_{xx}(\vecx) + \delta_{yy}(\vecx) - 2\delta_{zz}(\vecx)\right),\label{eqn_y20_mass_tensor}\\
\sigma_2 y_{21}^0(\vecx) & = & -\sqrt{15/2}\:\left(\delta_{xz}(\vecx) + \ci\delta_{yz}(\vecx)\right),\\
\sigma_2 y_{22}^0(\vecx) & = & +\sqrt{15/8}\:\left(\delta_{xx}(\vecx) - \delta_{yy}(\vecx) + 2\ci\delta_{xy}(\vecx)\right),\\
\sigma_2 y_{00}^1(\vecx) & = & +\left(\delta_{xx}(\vecx) + \delta_{yy}(\vecx) + \delta_{zz}(\vecx)\right)\label{eqn_y00_mass_tensor}.
\end{eqnarray}
The relation linking the tidal shear $\Psi_{\alpha\beta}$ to the $y_{\ell m}^n$-coefficients can be derived in complete analogy to eqns.~(\ref{eqn_y20_mass_tensor}) through (\ref{eqn_y00_mass_tensor}). The $y^n_{\ell m}$-coefficients of the tidal shear tensor field differ mainly by a factor of $\sigma_2/\sigma_0$ from those of the mass tensor field, apart from the trace of the tidal shear:
\begin{eqnarray}
\sigma_0 y_{20}^{-1}(\vecx) & = & -\sqrt{5/4}\:\left(\Psi_{xx}(\vecx) + \Psi_{yy}(\vecx) - 2\Psi_{zz}(\vecx)\right),\\
\sigma_0 y_{21}^{-1}(\vecx) & = & -\sqrt{15/2}\:\left(\Psi_{xz}(\vecx) + \ci\Psi_{yz}(\vecx)\right),\\
\sigma_0 y_{22}^{-1}(\vecx) & = & +\sqrt{15/8}\:\left(\Psi_{xx}(\vecx) - \Psi_{yy}(\vecx) + 2\ci\Psi_{xy}(\vecx)\right),\\
\sigma_0 y_{00}^0(\vecx) & = & +\left(\Psi_{xx}(\vecx) + \Psi_{yy}(\vecx) + \Psi_{zz}(\vecx)\right),
\end{eqnarray}
emphasising the difference in correlation length between the density field and the potential.

The $y_{\ell m}^n(\vecx)$-basis inherits its symmetry under complex conjugation from the spherical harmonics $Y_{\ell m}$:
\begin{equation}
y_{\ell m}^n(\vecx)^* = (-1)^m\: y_{l,-m}^n(\vecx),
\end{equation}
which will become important at the stage of inverting the relations given above. Similarly to the vector $\vecv$ containing the physical variables, the $y_{\ell m}^n(\vecx)$-coefficients can be arranged in a vector $\vecy$ by mapping the 3 indices $n$, $\ell$ and $m$ to a new index $i$. The physical $\vecv$-frame and the frame of the $\vecy$-values are related by a linear unitary transformation. For clarity, we abbreviate $\vecy\equiv\vecy(\vecx)$ and $\vecy^\prime\equiv\vecy(\vecx^\prime)$.

\subsection{Shape of the correlation matrices}\label{derivation_correlation_shape}
As demonstrated in the formalism proposed by  \citet{1995MNRAS.272..447R}, the correlation matrices needed in the Gaussian probability densities (eqns.~\ref{eqn_derivation_1pt_pdf} and \ref{eqn_derivation_2pt_pdf}) assume a particularly simple shape in the frame given by the $y_{\ell m}^n$-coefficients and can be expressed analytically in terms of moments of the dark matter power spectrum. The correlation matrix $\matrixy$ in this frame is defined as the expectation value 
\begin{equation}
Y_{ij}\equiv\bra(y, y^\prime)_i (y, y^\prime)_j^*\ket
\end{equation} 
of the products of the elements in the vector $(y,y^\prime)$ and can be split into two $15\times15$ submatrices: the auto-correlation matrix $\mathbfss{A}$, defined as $A_{ij}=\bra y_i y_j^*\ket$ and the cross-correlation matrix $\mathbfss{C}=\mathbfss{C}(r)$, given by $C_{ij}=\bra y_i y_j^{\prime*}\ket$, which depends on the distance $r$ between the two points $\vecx$ and $\vecx^\prime$,
\begin{equation}
\matrixy = \left(
\begin{array}{ll}
\mathbfss{A} & \mathbfss{C} \\ \mathbfss{C}^+ & \mathbfss{A}
\end{array}
\right),
\end{equation}
where $\mathbfss{C}^+$ is the Hermitean adjoint of $\mathbfss{C}$. The transformation between the $y_{\ell m}^n$-frame and the physical frame is given by the complex matrix $\matrixr$ acting on the vector $\bupsilon$ and resulting in the vector $\bmath{y}$, and by the matrix $\matrixs$, computing $(\bmath{y},\bmath{y}^\prime)$ from $\bmath{w}=(\vecv,\vecv^\prime)$,
\begin{equation}
\matrixs = \left(
\begin{array}{cc}
\matrixr & \mathbfss{0} \\ \mathbfss{0} & \matrixr
\end{array}
\right).
\end{equation}
The matrices $\matrixr$ and $\matrixs$ can be constructed from the relations between the $y_{\ell m}^n(\vecx)$-coefficients and the physical variables $\delta(\vecx)$, $\delta_\alpha(\vecx)$, $\delta_{\alpha\beta}(\vecx)$ and $\Psi_{\alpha\beta}(\vecx)$ compiled in Sect.~\ref{derivation_spherical_coords}.

\subsubsection{Auto-correlation matrix}
The auto-correlation matrix $\mathbfss{A}$ in the $y_{\ell m}^n$-frame is defined by:
\begin{equation}
\mathbfss{A} = 
A_{\ell\lprime m\mprime}^{n\nprime} = 
\bra y_{\ell m}^n(\vecx) y_{\lprime\mprime}^{\nprime}(\vecx)^*\ket.
\end{equation}
Inserting the Fourier expansion of the variables, replacing the variance of the density field $\bra\delta(\veck)\delta(\veck^\prime)^*\ket$ with the matter power spectrum $P(K)$, and using the orthogonality relation of the spherical harmonics $Y_{\ell m}(\hat{k})$,
\begin{equation}
\int\dd\Omega\:Y_{\ell_1 m_1}(\hat{k}) Y_{\ell_2 m_2}(\hat{k})^* = \delta_{\ell_1\ell_2}\delta_{m_1m_2},
\end{equation}
yields for the auto-correlation matrix:
\begin{equation}
A_{\ell\lprime m\mprime}^{n\nprime} 
= A_{\ell m}^{n\nprime}\delta_{\ell\lprime}\delta_{m\mprime} 
= (-1)^{n-\nprime}
\frac{\sigma^2_{\ell+n+\nprime}}{\sigma_{\ell+2n}\sigma_{\ell+2n^\prime}}\delta_{\ell\lprime}\delta_{m\mprime},
\end{equation}
where the definition of the $\sigma_j$-coefficients in eqn.~(\ref{eqn_derivation_sigmaj}) was used for substituting the $(2j+2)^\mathrm{th}$ moments of the power spectrum $P(k)$. The structure of the matrix $\mathbfss{A}$ is remarkably simple: It is diagonal in the indices $\ell$ and $m$ and the sign of its entirely real entries is determined by whether $n-\nprime$ is an even or odd number.

\subsubsection{Cross-correlation matrix}
The cross-correlation matrix $\mathbfss{C}(r)$ is defined analogously,
\begin{equation}
\mathbfss{C}(r) = 
C_{\ell\lprime m\mprime}^{n\nprime}(r) = 
\bra y_{\ell m}^{n}(\vecx) y_{\lprime\mprime}^{\nprime}(\vecx^\prime)^*\ket.
\end{equation}
The steps in simplifying this expression consist in inserting the definition of the $y_{\ell m}^n(\vecx)$-coefficients, in replacing the variance $\bra\delta(\veck)\delta(\veck^\prime)^*\ket$ with the matter power spectrum $P(k)$ and in expanding the Fourier wave $\exp(\ci\veck\vecr)$, $\vecr\equiv\vecx-\vecx^\prime$, by virtue of Rayleigh's formula,
\begin{equation}
\exp(\ci\veck\vecr) = 
4\pi\sum_{L=0}^\infty \ci^L j_L(kr) \sum_{M=-L}^{+L} Y_{LM}(\hat{r})^* Y_{LM}(\hat{k}).
\label{eqn_rayleigh_expansion}
\end{equation}
The integration over the three spherical harmonics $Y_{\ell m}(\hat{k})$ can be simplified by inserting the definition of the Wigner-$3j$ symbols \citep{1962qume.book.....M, 1988AmJPh..56..958A},
\begin{displaymath}
\int\dd\Omega\: Y_{\ell_1 m_1}(\hat{k}) Y_{\ell_2 m_2}(\hat{k})^* Y_{\ell_3 m_3}(\hat{k}) = 
\end{displaymath}
\begin{equation}
(-1)^{m_2}\sqrt{\frac{\Pi_{i=1}^3 (2\ell_i+1)}{4\pi}}
\left(
\begin{array}{ccc}
\ell_1 & \ell_2 & \ell_3 \\ 0 & 0 & 0
\end{array}
\right)
\left(
\begin{array}{ccc}
\ell_1 & \ell_2 & \ell_3 \\ m_1 & -m_2 & m_3
\end{array}
\label{eqn_def_wigner_3j}
\right).
\end{equation}
Further reduction is reached by taking advantage of the fact that both peaks are assumed to lie on the $z$-axis,
\begin{equation}
Y_{LM}(\hat{r}) = \delta_{M0}\sqrt{\frac{2L+1}{4\pi}},
\end{equation}
which yields the final form of the cross-correlation matrix $\mathbfss{C}(r)$:
\begin{displaymath}
C_{\ell\lprime m\mprime}^{n\nprime}(r)
= \delta_{m\mprime}\frac{(-1)^{m+n-\nprime}}{\sigma_{\ell+2n}\sigma_{\lprime+2\nprime}}
\sum_{L=\left|\ell-\lprime\right|}^{\ell+\lprime} (2L+1)\ci^{L+\ell-\lprime} K_{L,\ell+\lprime+2(n+\nprime+1)}(r)\nonumber
\end{displaymath}
\begin{equation}
\quad\times\sqrt{(2\ell+1)(2\lprime+1)}
\left(
\begin{array}{ccc}
\ell & \lprime & L \\ 0 & 0 & 0
\end{array}
\right)
\left(
\begin{array}{ccc}
\ell & \lprime & L \\ m & -m & 0
\end{array}
\right),
\end{equation}
where the $k^m j_\ell(kr)$-weighted spectral moments are abbreviated with $K_{\ell m}(r)$,
\begin{equation}
K_{\ell m}(r) = \frac{1}{2\pi^2}\int\dd k\:k^m j_\ell(kr) P(k).
\end{equation}
$j_\ell(kr)$ are the spherical Bessel functions of the first kind \citep{1988AmJPh..56..958A}. The $C^{n\nprime}_{\ell\lprime m\mprime}$-coefficients are always real: The Wigner $3j$-symbols are unequal to zero if $L+\ell-\lprime$ is even, in which case $\ci^{L+\ell-\lprime}$ is a real number. Furthermore, the summation over $L$ can be restricted to the range $|\ell-\lprime|\leq L\leq\ell+\lprime$ due to the triangle condition applied to the Wigner-$3j$ symbols. The cross-correlation matrix $\mathbfss{C}$ can be brought to block diagonal shape by suitable arrangement of the $y^n_{\ell m}$-coefficients in the vector $\vecy$, more specifically, by grouping coefficients with constant value of $m$ and increasing the modulus of $m$ with increasing index $i$.

\begin{figure}
\resizebox{\hsize}{!}{\includegraphics{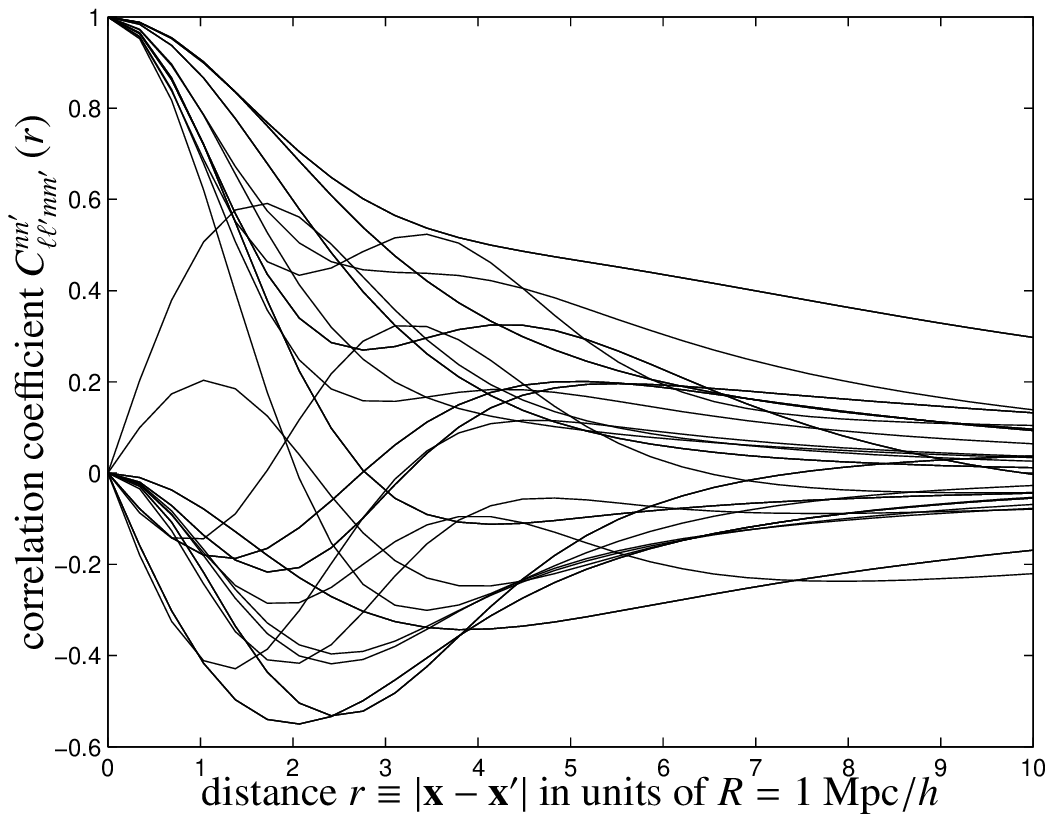}}
\caption{Correlation coefficients $C^{n\nprime}_{\ell\lprime m\mprime}(r)=\bra y^{n}_{\ell m}(\mathbf{x}) y^{\nprime}_{\lprime \mprime}(\mathbf{x}^\prime)^*\ket$ as a function of distance $r\equiv\left|\mathbf{x}-\mathbf{x}^\prime\right|$ in units of the filter scale $R$, transformed into a frame, where the auto-correlation matrix $A^{n\nprime}_{\ell\lprime m\mprime}$ is equal to the {\em unit matrix}. The smoothing scale $R$ was set to $R=1~\mathrm{Mpc}/h$ in order to represent galaxies.}
\label{fig_covariance_cframe}
\end{figure}

In contrast to the constant values in the matrix $\mathbfss{A}$, the entries of the matrix $\mathbfss{C}$ depend on the distance $r=\left|\vecx-\vecx^\prime\right|$ of the two points $\vecx$ and $\vecx^\prime$. The symmetry of the entries of $\mathbfss{C}$ under interchange of the points $\vecx$ and $\vecx^\prime$ is given by the relation
\begin{equation}
\bra y^n_{\ell m}(\vecx) y^{\nprime}_{\lprime\mprime}(\vecx^\prime)^*\ket = 
(-1)^{\ell-\lprime}\bra y^n_{\ell m}(\vecx^\prime)^* y^{\nprime}_{\lprime\mprime}(\vecx)\ket.
\end{equation}
Typical correlation coefficients $C^{n\nprime}_{\ell\lprime m\mprime}$ as functions of separation $r$ are depicted in Fig.~\ref{fig_covariance_cframe}. The smoothing scale $R$ (c.f. eqn.~\ref{eqn_smoothed_power_spectrum}) has been set to $R=1~\mathrm{Mpc}/h$ and the density threshold was chosen as $\nu=2$, in order to represent galaxies. With the  choice of $\Omega_m$, the smoothing of the power spectrum at scale $R$ corresponds to a mass scale of $M_\mathrm{scale}=\frac{4\pi}{3}\rho_\mathrm{crit}\Omega_m R^3\simeq3.1\times10^{11}M_\odot/h$. For illustration purposes, the covariance matrix $C^{n\nprime}_{\ell\lprime m\mprime}$ has been transformed to a frame, where the auto-correlations $A^{n\nprime}_{\ell\lprime m\mprime}$ are diagonal and normalised to unity.

\begin{figure}
\resizebox{\hsize}{!}{\includegraphics{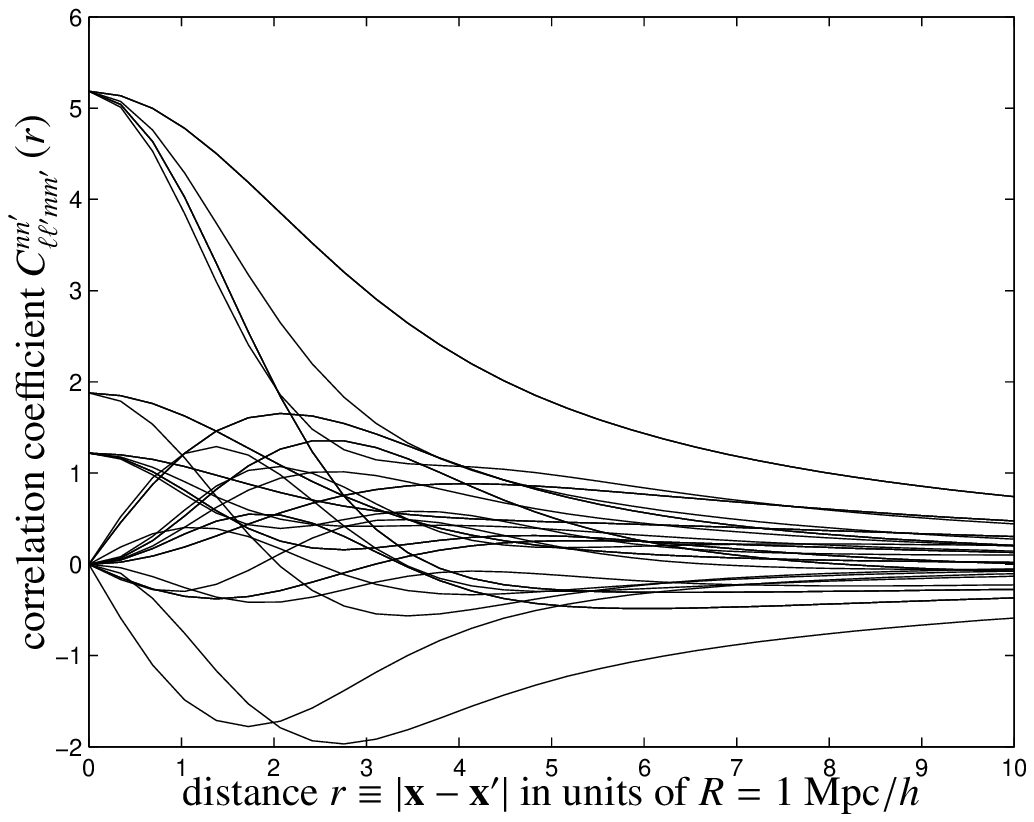}}
\caption{Correlation coefficients $C^{n\nprime}_{\ell\lprime m\mprime}(r)=\bra y^{n}_{\ell m}(\mathbf{x}) y^{\nprime}_{\lprime \mprime}(\mathbf{x}^\prime)^*\ket$ as a function of distance $r\equiv\left|\mathbf{x}-\mathbf{x}^\prime\right|$ in units of the filter scale $R$, transformed into a frame, where the auto-correlation matrix $A^{n\nprime}_{\ell\lprime m\mprime}$ is {\em diagonal}. The smoothing scale $R$ was set to $R=1~\mathrm{Mpc}/h$ in order to represent galaxies.}
\label{fig_covariance_jframe}
\end{figure}

Fig.~\ref{fig_covariance_jframe} shows the distance dependence of the entries of $C^{n\nprime}_{\ell\lprime m\mprime}(r)$ in a frame, where the auto-correlation matrix $A^{n\nprime}_{\ell\lprime m\mprime}$ is diagonal instead of equal to the unit matrix. Notable differences to $C^{n\nprime}_{\ell\lprime m\mprime}$ in this frame compared to the frame where to auto-correlations are unity include the fact that $C^{n\nprime}_{\ell\lprime m\mprime}$ indicates the elongation of the isoprobability contours, which assume semi-axis ratios of $\simeq5$ and a less oscillatory behaviour.

\subsection{Ansatz for the angular momentum correlations}\label{derivation_ll_correlation}
The correlation function of the angular momenta in the large-scale structure follows from integrating out the 30d Gaussian probability density $p(\vecw)\dd\vecw$, constraint to peak regions in the fluctuating density field. The regions to which the integration is restricted are required to exceed a threshold $\nu$ in density, $\delta(\vecx)>\nu\sigma_0$, to be of vanishing density gradient, $\delta_\alpha(\vecx)=0$, and of negative curvature, $\delta_{\alpha\beta}(\vecx)<0$.

The number density of maxima in the density field is modeled by a point-process \citep{1986ApJ...304...15B, 1995MNRAS.272..447R, 1999MNRAS.310.1062H}:
\begin{equation}
n_\mathrm{peak} = \sum_i\delta_D^3(\vecx-\vecx_i).
\end{equation}
Close to the maximum $i$ at $\vecx_i$ a Taylor expansion of the density gradient $\delta_\alpha(\vecx)$ is applicable:
\begin{equation}
\delta_\alpha(\vecx) = \sum_\beta\delta_{\alpha\beta}(\vecx_i)(\vecx-\vecx_i)_\beta.
\end{equation}
With this expansion one obtains for the peak density:
\begin{equation}
n_\mathrm{peak} 
= \sum_i\delta_D^3\left(\delta_{\alpha\beta}^{-1}(\vecx_i)\delta_\alpha\right) 
= \left|\determinant\:\delta_{\alpha\beta}\right|\:\delta_D^3(\delta_\alpha).
\end{equation}
The relation takes a simpler shape when considering the eigensystem of the mass tensor $-\delta_{\alpha\beta}$: Being a symmetric tensor, it has the three real eigenvalues $\lambda_i$, $i=1\ldots3$ which allows to replace the determinant by the invariant quantity $\left|\lambda_1\lambda_2\lambda_3\right|$.

The constraints can be combined in a mask $\mathcal{C}(\vecv)$, which is defined as a function on the vector $\vecv$ containing the derivates of the Gaussian random field under consideration: Peaks in the density field are defined as points with amplitudes in units of the variance $\sigma_0^2=\bra\delta^2\ket$ exceeding a certain threshold $\nu$ and exhibiting a vanishing gradient $\delta_\alpha$ as well as negative curvature $\delta_{\alpha\beta}$:
\begin{equation}
\mathcal{C}(\vecv) = 
\delta^3_D\left[\delta_\alpha(\vecx)\right]\:\left|\lambda_1\lambda_2\lambda_3\right|\:\prod_i\Theta(\lambda_i)\:
\Theta\left[\delta(\vecx)-\sigma_0\nu\right].
\end{equation}
$\Theta(x)$ denotes the Heaviside step-function. The peak density $n_\mathrm{peak}$, i.e. the expecation value for the number density of peaks in the fluctuating density field $\delta$ which exceed a threshold $\nu\sigma_0$ can then be derived from the multivariate Gaussian random process $p(\vecv)\dd\vecv$,
\begin{equation}
n_\mathrm{peak} =\int\dd\vecv\: p(\vecv)\:\mathcal{C}(\vecv),
\label{eqn_peak_density}
\end{equation}
which corresponds to the integral of the differential peak density $n_\mathrm{peak}(\nu)\dd\nu$ defined by \citet{1986ApJ...304...15B}.

In analogy, the expectation value of the angular momentum restricted to peak regions in the density field can be obtained with:
\begin{equation}
\bra\vecl\ket = \frac{1}{n_\mathrm{peak}(>\nu)}\int\dd\vecv\: p(\vecv)\: \mathcal{C}(\vecv) \vecl(\vecv),
\label{derivation_eqn_l_expectation}
\end{equation}
and the variance of the angular momentum field can be computed using:
\begin{equation}
\bra\vecl^2\ket = \frac{1}{n_\mathrm{peak}(>\nu)}\int\dd\vecv\: p(\vecv)\: \mathcal{C}(\vecv) \vecl^2(\vecv).
\label{eqn_l_1pt_variance}
\end{equation}
In both formulae, the normalisation factor $n_\mathrm{peak}^{-1}$ accounts for the discreteness of the measured quantity.

Generalisation of the above relations to include a second peak results in the correlation function $\bra L_\alpha(\vecx) L_{\alpha^\prime}(\vecx^\prime)\ket$ of the angular momenta, with the Gaussian probability density $p(\vecw)\dd\vecw=p(\vecv,\vecv^\prime)\dd\vecv\dd\vecv^\prime$:
\begin{equation}
\bra L_\alpha(\vecx) L_{\alpha^\prime}(\vecx^\prime)\ket = 
\label{eqn_ll_corr}
\end{equation}
\begin{displaymath}
\quad\frac{1}{n_\mathrm{peak}^2(>\nu)}
\int\dd\vecv\mathcal{C}(\vecv)\:\int\dd\vecv^\prime\mathcal{C}(\vecv^\prime)\: 
L_\alpha(\vecv) L_{\alpha^\prime}(\vecv^\prime) p(\vecv,\vecv^\prime).
\end{displaymath}
In general, the thresholds $\nu$, $\nu^\prime$ imposed on the peaks are equal. As derived in Sect.~\ref{derivation_zeldovich_shear}, the angular momentum $L_\alpha$ depends on the product of the inertia tensor $I_{\beta\sigma}$ and the tidal shear $\Psi_{\sigma\gamma}$:
\begin{equation}
L_\alpha 
= a^2 \dot{D}_+ \epsilon_{\alpha\beta\gamma}\sum_\sigma I_{\beta\sigma} \Psi_{\sigma\gamma}
= a^2 \dot{D}_+ \epsilon_{\alpha\beta\gamma}X_{\beta\gamma}, 
\end{equation}
if the acquisition of angular momentum of a protogalactic object is described in the Zel'dovich approximation. For convenience, we introduce the matrix $\matrixx$ with the components:
\begin{equation}
X_{\beta\gamma}(\vecx) = \sum_\sigma I_{\beta_\sigma}(\vecx)\Psi_{\sigma\gamma}(\vecx).
\end{equation}
Then, the correlation of the angular momentum components becomes:
\begin{equation}
\bra L_\alpha(\vecx) L_{\alpha^\prime}(\vecx^\prime)\ket = a^4 \dot{D}_+^2
\epsilon_{\alpha\beta\gamma}\epsilon_{\alpha^\prime\beta^\prime\gamma^\prime}
\bra X_{\beta\gamma}(\vecx) X_{\beta^\prime\gamma^\prime}(\vecx^\prime)\ket.
\label{eqn_derivation_ll_corr_matrix}
\end{equation}
In the next step we replace the 1d variance of the components $L_\alpha$ in the correlation function by the 3d variance of the full vector $\vecl$ by taking the trace of eqn.~(\ref{eqn_derivation_ll_corr_matrix}),
\begin{equation}
C_L(r)
\equiv\trace\bra L_\alpha(\vecx) L_{\alpha^\prime}(\vecx^\prime)\ket
= \bra L_\alpha(\vecx) L_{\alpha}(\vecx^\prime)\ket,
\label{eqn_trace_ll_corr}
\end{equation}
which has the advantage of being a coordinate-frame independent quantity and allows the usage of the relation $\epsilon_{\alpha\beta\gamma}\epsilon_{\alpha\beta^\prime\gamma^\prime} = 3\left(\delta_{\beta\beta^\prime}\delta_{\gamma\gamma^\prime} - \delta_{\beta\gamma^\prime}\delta_{\beta^\prime\gamma}\right)$ for reducing the product of the two $\epsilon_{\alpha\beta\gamma}$-symbols:
\begin{equation}
\bra L_\alpha(\vecx) L_\alpha(\vecx^\prime)\ket = 3a^4 \dot{D}_+^2
\left[
\bra X_{\beta\gamma}(\vecx^\prime) X_{\beta\gamma}(\vecx) \ket - \bra X_{\beta\gamma}(\vecx) X_{\gamma\beta}(\vecx^\prime)\ket
\right],
\end{equation}
where the order of the indices in the last term is interchanged. In matrix notation, the correlation function $C_L(r)$ reads:
\begin{equation}
C_L(r)
= \trace\bra \vecl(\vecx) \vecl^t(\vecx^\prime)\ket 
= 3a^4 \dot{D}_+^2 \:\trace
\left[\bra\matrixx(\vecx)\matrixx^t(\vecx^\prime)\ket -
\bra\matrixx(\vecx^\prime)\matrixx(\vecx)\ket\right],
\label{eqn_derivation_ll_ttranspose}
\end{equation}
which is non-vanishing for general asymmetric matrices $X_{\beta\gamma}$ due to the matrix transposition in the last term. Finally, one can express the angular momentum correlation arising from the Gaussian probability density in the natural variable $\matrixx$, $X_{\beta\gamma}=I_{\beta\sigma}\Psi_{\sigma\gamma}$:
\begin{equation}
\trace\bra \vecl(\vecx)\vecl^t(\vecx^\prime)\ket = 3\frac{a^4 \dot{D}_+^2}{n_\mathrm{peak}^2(>\nu)}\times
\label{eqn_trace_ll_w_integral}
\end{equation}
\begin{displaymath}
\quad
\int\dd\vecv\mathcal{C}(\vecv)\:\int\dd\vecv^\prime\mathcal{C}(\vecv^\prime)\:
\trace\left[\matrixx(\vecv)\matrixx^t(\vecv^\prime) - \matrixx(\vecv^\prime)\matrixx(\vecv)\right] p(\vecv,\vecv^\prime).
\end{displaymath}
In the calculation outlined above we aim to avoid a decomposition of the tidal shear and inertia correlations according to
\begin{equation}
\bra\matrixi(\vecx)\bmath{\Psi}(\vecx)\:\matrixi(\vecx^\prime)\bmath{\Psi}(\vecx^\prime)\ket =
P(\matrixi|\bmath{\Psi})\: P(\matrixi|\bmath{\Psi})\:
\bra\bmath{\Psi}(\vecx)\:\bmath{\Psi}(\vecx^\prime)\ket,
\end{equation}
which uses the difference in correlation lengths of the $\bmath{\Psi}$ and $\matrixi$-fields and is valid on scales on which the $\matrixi$-tensors are uncorrelated.

\subsection{Misalignment of the shear and inertia eigensystems}\label{derivation_misalignment}
The tensor $\matrixx=\mathbfss{I}\bmath{\Psi}$ can be decomposed according to $\matrixx = \matrixx^+ + \matrixx^-$ into an antisymmetric contribution $\matrixx^-$, defined via the commutator $\left[\mathbfss{I},\bmath{\Psi}\right]$,
\begin{equation}
\matrixx^- \equiv \frac{1}{2}\left[\mathbfss{I},\bmath{\Psi}\right], \quad
X_{\beta\gamma}^- = \frac{1}{2}\sum_\sigma\left(I_{\beta\sigma}\Psi_{\sigma\gamma} - \Psi_{\beta\sigma}I_{\sigma\gamma}\right),
\end{equation}
with the symmetry $(\matrixx^-)^t = \frac{1}{2}(\mathbfss{I}\bmath{\Psi}-\bmath{\Psi}\mathbfss{I})^t = \frac{1}{2}(\bmath{\Psi}\mathbfss{I} - \mathbfss{I}\bmath{\Psi}) = - \matrixx^-$ under matrix transposition ($\mathbfss{I}$ and $\bmath{\Psi}$ are symmetric matrices) and into the corresponding symmetric matrix $\matrixx^+$ by using the anticommutator $\left\{\mathbfss{I},\bmath{\Psi}\right\}$ between inertia $\mathbfss{I}$ and tidal shear $\bmath{\Psi}$:
\begin{equation}
\matrixx^+ 
\equiv \frac{1}{2}\left\{\mathbfss{I},\bmath{\Psi}\right\},\quad
X_{\beta\gamma}^+ = \frac{1}{2}\sum_\sigma\left(I_{\beta\sigma}\Psi_{\sigma\gamma} + \Psi_{\beta\sigma}I_{\sigma\gamma}\right),
\end{equation}
with $(\matrixx^+)^t = +\matrixx^+$. In the derivation of the angular momentum $\vecl$,
\begin{equation}
L_\alpha 
= a^2 \dot{D}_+ \epsilon_{\alpha\beta\gamma}\sum_\sigma I_{\beta\sigma}\Psi_{\sigma\gamma}
= a^2 \dot{D}_+ \epsilon_{\alpha\beta\gamma} X_{\beta\gamma},
\end{equation}
the permutation symbol $\epsilon_{\alpha\beta\gamma}$ picks out the antisymmetric contribution $\matrixx^-$, by virtue of $\epsilon_{\alpha\beta\gamma}(X_{\beta\gamma}^+ + X_{\beta\gamma}^-) = X_{\beta\gamma}^-$, because the contraction of the symmetric tensor $\matrixx^+$ with the antisymmetric permutation symbol vanishes, $\epsilon_{\alpha\beta\gamma} X_{\beta\gamma}^+=0$. 

Hence, the protogalactic objects will only acquire angular momentum if the commutator $\matrixx^-$ between the inertia and the tidal shear is non-zero, which means that the inertia and shear tensors are not supposed to be simultaneously diagonisable, i.e. they are not allowed to have a common eigensystem. In order to capture this mechanism, \citet{2000ApJ...532L...5L} and \citet{2001ApJ...559..552C} have used an effective, parameterised description of the average misalignment of the shear and inertia eigensystems, gauged with numerical $n$-body data. 

In the correlation function of the angular momenta $C_L(r)=\trace\bra\vecl^t(\vecx)\vecl(\vecx^\prime)\ket$ (c.f. eqn.~\ref{eqn_derivation_ll_ttranspose}), the dependence of $\vecl$ on the commutator $\matrixx^-$ translates into the asymmetric quadratic form $\bra\matrixx(\vecx)\matrixx(\vecx^\prime)-\matrixx(\vecx)\matrixx^t(\vecx^\prime)\ket$ with the matrix transpose in the second term carrying the signal: A common eigensystem of the inertia and shear tensors, being both symmetric matrices, would have the consequence that $\matrixx$ would be a symmetric matrix, $\matrixx=\matrixx^+=\matrixx^t$, and the correlation function $C_L(r)$ would vanish. 

In contrast, the signal is maximised, if the shear and inertia eigensystems are unaligned, i.e. if $\matrixx$ is purely antisymmetric, $\matrixx=\matrixx^-$ and $\matrixx^+=0$. In that case $\matrixx^t=(\matrixx^-)^t=-\matrixx^-=-\matrixx$ and the angular momentum auto-correlation function $C_L^\mathrm{max}(r)$ is truly quadratic:
\begin{equation}
C_L^\mathrm{max}(r) = 6\:\frac{a^4\dot{D}_+^2}{n_\mathrm{peak}^2(>\nu)}
\int\dd\vecv\mathcal{C}(\vecv) \int\dd\vecv^\prime\mathcal{C}(\vecv^\prime) \:
\trace\left[\matrixx(\vecv)\matrixx^t(\vecv^\prime)\right].
\end{equation}
The corresponding 1-point variance would be given by:
\begin{equation}
\bra\vecl^2_\mathrm{max}\ket = \frac{1}{n_\mathrm{peak}(>\nu)}\int\dd\vecv\: p(\vecv)\: \mathcal{C}(\vecv) \vecl^2_\mathrm{max}(\vecv),
\label{eqn_lmax_1pt_variance}
\end{equation}
which illustrates the effect of partial alignment of the shear and inertia eigensystems, reducing the variance compared to the case where the shear and inertia eigensystems are maximally misaligned.

The symmetric contribution $\matrixx^+$, which measures the degree of alignment of the inertia and shear eigensystems, causes an anisotropic deformation of the protogalactic region during the course of linear structure formation prior to gravitational collapse. Consequently, the determination of ellipticity distributions is likely to be affected even in the stage of linear structure formation, and predictions of triaxiality based on peak shapes in Gaussian random fields \citep{1986ApJ...304...15B} could be refined using an adaptation of the formalism outlined above.

\subsection{Symmetry of the cancellation mechanism}\label{derivation_cancellation}
A possible objection concerning the symmetry of the correlation function $C_L(r)$,
\begin{equation}
C_L(r) \propto \left\langle 
\trace\left(\matrixx(\vecx)\matrixx^t(\vecx^\prime) - \matrixx(\vecx^\prime)\matrixx(\vecx)\right)
\right\rangle,
\end{equation}
might be that the mechanism comparing the alignments of the shear and inertia eigensystems outlined above is only present in the matrix $\matrixx(\vecx^\prime)$ at position $\vecx^\prime$, which appears transposed in the second term, but not in the matrix $\matrixx(\vecx)$ at position $\vecx$. The expression, however, can be reformulated using $\trace(\matrixx(\vecx^\prime)\matrixx(\vecx)-\matrixx(\vecx)\matrixx(\vecx^\prime)^t)=\trace(\matrixx(\vecx^\prime)\matrixx(\vecx)-(\matrixx(\vecx)\matrixx(\vecx^\prime)^t)^t)=\trace(\matrixx(\vecx)\matrixx(\vecx^\prime)-\matrixx(\vecx^\prime)\matrixx(\vecx)^t)$, using the properties of the trace $\trace(\amatrix+\bmatrix)=\trace(\amatrix)+\trace(\bmatrix)$, $\trace(\amatrix\bmatrix)=\trace(\bmatrix\amatrix)$ and $\trace(\amatrix^t)=\trace(\amatrix)$, as well as $(\amatrix^t)^t=\amatrix$ and $(\amatrix\bmatrix)^t=\bmatrix^t \amatrix^t$ such that the mechanism is present in the first tensor as well.

Another objection might be that the correlation function $C_L(r)$ vanishes if one of the matrices $\matrixx$ is symmetric. A symmetric shape of the matrix $\matrixx$, however, never occurs in the angular momentum build-up, because of the fact that symmetric matrices form a group under matrix multiplication: A symmetric matrix $\matrixx$ could only have emerged from symmetric matrices $\mathbfss{I}$ and $\bmath{\Psi}$ with a common eigensystem due to a vanishing commutator $\left[\mathbfss{I},\bmath{\Psi}\right]$, making it impossible for the halo to acquire angular momentum as discussed earlier, and consequently, the correlation function has to be zero.

\subsection{Numerics of the constraint Gaussian integration}\label{derivation_montecarlo}
The covariance matrix $\matrixy$ in the frame given by the $y^n_{\ell m}$-coefficients is transformed to the physical frame yielding the correlation matrix $\matrixw$. Then, we determine numerically the correlation functions $C_L(r)=\trace\bra \vecl^t(\vecx)\vecl(\vecx^\prime)\ket$ of the angular momentum $\vecl$ as a function of distance $r=\left|\vecx-\vecx^\prime\right|$, by carrying out the integration over the multivariate Gaussian probability density $p(\vecw)\dd\vecw$, subjected to being constraint to the peaks in the density field by the mask $\mathcal{C}(\vecw)=\mathcal{C}(\vecv)\mathcal{C}(\vecv^\prime)$. The inertia $I_{\alpha\beta}$ of an object forming at a peak in the density field is consistently derived from the local curvature $\partial_\alpha\partial_\beta\delta(\vecx)$ of the density field at the peak.

Due to the high dimensionality the numerical constraint integration is a difficult and time consuming task. This task is most efficiently addressed by exploiting our prior knowledge of the underlying probability density function: Since the distribution is just -- tough very high dimensional -- Gaussian, it is advantageous to sample the integral of eqn.~(\ref{eqn_trace_ll_w_integral}) directly instead of relying blindly on a common Monte-Carlo scheme.
Generating samples which follow the distribution given in eqn.~(\ref{eqn_derivation_1pt_pdf}) and eqn.~(\ref{eqn_derivation_2pt_pdf}) respectively is straightforward since this can be mapped onto the generation of unit Gaussian variates, for example via the Cholesky decomposition of the corresponding covariance matrix. Gaussian (unit) variates, however, can be obtained from a variety of very fast and efficient random number generators, e.g. the ziggurat method.

Following the strategy of reducing the numerical integration of eqn.~(\ref{eqn_trace_ll_w_integral}) to a direct sampling process which only requires the generation of unit Gaussian variates is the most important simplification and acceleration in comparison to the use of standard Monte-Carlo techniques. In addition, our method allows to carry out the sampling in the physical frame where the constraints \(C\) can be evaluated most easily. Further acceleration can be achieved by using simple linear algebra in order to minimize the number of diagonalisations of the mass tensor $M_{\alpha\beta}=\partial_\alpha\partial_\beta\delta$. Instead of checking whether all eigenvalues of the mass tensor are negative we check its negative definiteness by combining the invariants $\trace(M^n)$, $n=1,2$ and the determinant $\mathrm{det}(M)$. Finally, it is important to note that the actual dimensionality of the integration is reduced by the requirement of being at a peak in the density field, i.e. \(\delta_D(\nabla \delta(\mathbf x ) )\). Thus, in our sampling process the derivatives of the density field are no degree of freedom but rather fixed (namely to be zero). This requires to adjust the overall normalisation of our constrained Gaussian sampling process by an appropriate factor, i.e. we have to use the normalisation of the unconstrained Gaussian distribution.

In order to investigate the performance and reliability of our sampling strategy used to evaluate integrals like those of eqn.~(\ref{eqn_l_1pt_variance}) and eqn.~(\ref{eqn_trace_ll_w_integral}) we computed in a first step the mean peak density given in eqn.~(\ref{eqn_peak_density}) and compared our result with the analytical solution derived by \citet{1986ApJ...304...15B}. Increasing the number of samples we could reproduce their analytical result to arbitrary accuracy. We could also recover the limiting case \(\nu\rightarrow-\infty\). To provide an error estimate for our results of the angular momentum correlation functions given in the next section we carried out every sampling process several times seeding the random number generator differently and computed the resulting \(1 \sigma\) error. For all numerical calculations we assured that the number of accepted samples is of the order of \(10^5\) for each step in distance.

\section{Results}\label{result_l_correlation}
For visualisation, we define the specific angular momentum
\begin{equation}
\tilde{\bmath{L}} = \frac{\vecl}{H_0 M_\mathrm{scale}},
\end{equation}
normalised by the mass scale $M_\mathrm{scale}=\frac{4\pi}{3}\rho_\mathrm{crit}\Omega_m R^3$ and divide out the Hubble-constant $H_0$ defining the cosmological time scale such that the resulting quantity has units of a squared length scale and depends only on a single fundamental unit, for which we choose $\mathrm{Mpc}/h$. 

\begin{figure}
\resizebox{\hsize}{!}{\includegraphics{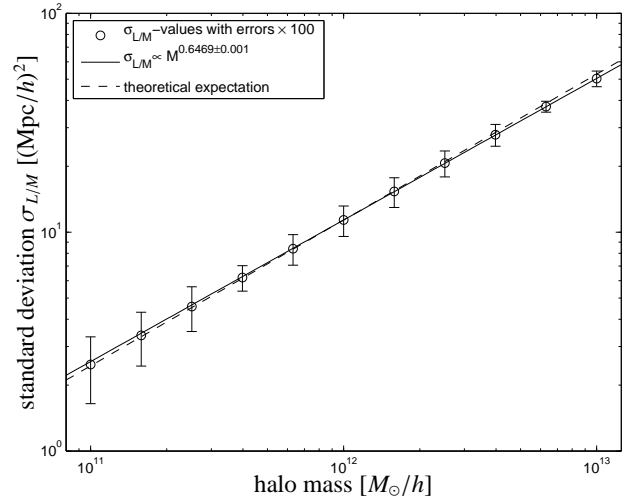}}
\caption{Standard deviation $\sigma_{L/M}$ of the specific angular momentum as a function of halo mass, as obtained from numerical sampling (circles) including error bars (artificially enlarged by a factor of 100), along with a power-law fit to the data (solid line) and the theoretical expectation $\sigma_{L/M}\propto M^{2/3}$ (dashed line).}
\label{fig_sigma_scaling}
\end{figure}

As shown in \citet{1969ApJ...155..393P} and \citet{1988MNRAS.232..339H}, the standard deviation $\sigma_{L/M}$ of this quantity scales $\propto M^{2/3}$, which is a valuable check for our numerical sampling code. In fact, as shown in Fig.~\ref{fig_sigma_scaling}, we recover this behaviour to a high degree of accuracy over two orders of magnitude in halo mass. A fit to the data points including their sampling error yields
\begin{equation}
\frac{\sigma_{L/M}}{\sigma_{L/M,0}} = \left(\frac{M}{M_0}\right)^{0.6469\pm0.001},
\end{equation}
with $\sigma_{L/M} = 11.362\sigma_0\:(\mathrm{Mpc}/h)^2$ and $M_0 = 10^{12} M_\odot/h$, within 3\% of the theoretical value. For convenience, we have divided $L/M$ in Fig.~\ref{fig_sigma_scaling} by $\sigma_0$. Physically, the $1\sigma$ value of $L/M$ corresponds to a homogeneous sphere with the radius $1~\mathrm{Mpc}/h$ and a mass of $10^{12}M_\odot/h$ revolving once every $\sim10^9$ years.

\begin{figure}
\resizebox{\hsize}{!}{\includegraphics{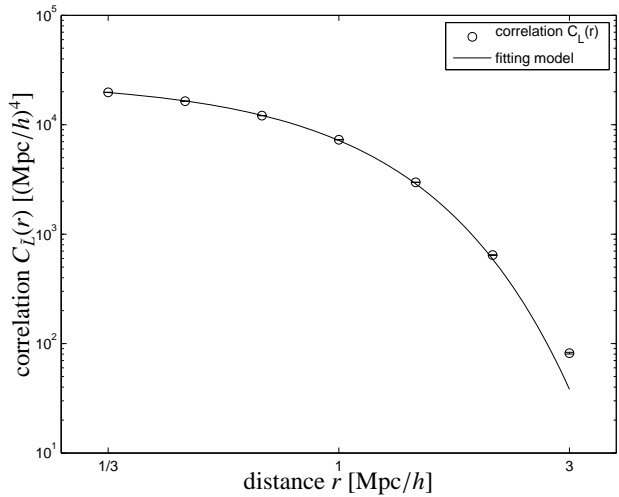}}
\caption{Angular momentum correlation function $C_{\tilde{L}}(r)$ for a Milky Way-sized halo of $10^{12}M_\odot/h$ (circles) including sampling errors, together with an empirical fitting formula of the form $C_{\tilde L}(r)\propto\exp(-[r/r_0]^\beta)$ (solid line).}
\label{fig_ll_spectrum}
\end{figure}

The correlation function $C_{\tilde{L}}(r)$ of the specific angular momentum is depicted in Fig.~\ref{fig_ll_spectrum} for a mass scale of $10^{12}M_\odot/h$ corresponding to a Milky Way-sized halo. It is decreasing rapidly after a distance of $\sim 1~\mathrm{Mpc}/h$ corresponding to the correlation length of the angular momentum field and assumes the asymptotic value of $C_{\tilde L}(r)\rightarrow \sigma_{\tilde L}^2/n_\mathrm{peak}$ for $r\rightarrow 0$ in fulfilment of the Cauchy-Schwarz inequality with our values for the peak density. An empirical fit to the spectrum is given by
\begin{equation}
C_{\tilde L}(r) = a\exp\left(-\left[r/r_0\right]^\beta\right)
\end{equation}
with $a=(25010\pm165)\sigma_0^2~(\mathrm{Mpc}/h)^4$, $r_0 = (0.8628\pm0.008)~\mathrm{Mpc}/h$ (both error bounds correspond to $1\sigma$) and $\beta=3/2$. In addition, we find the angular momenta of haloes to be positively correlated, i.e. angular momentum vectors tend to be aligned in a parallel way, because the form of our correlation function would be able to distinguish between parallel and antiparallel alignment.

\section{Summary}\label{sect_summary}
In this paper, we recompute the angular momentum correlation function arising from tidal torquing in an improved statistical model, which is based on a peak-restricted Gaussian random process.
\begin{enumerate}
\item{Dark energy influences angular momentum aquisition by the time derivative $H(a)\dd D_+/\dd a$ of the growth function $D_+(a)$. Angular momentum build-up in dark energy models is significantly slower compared to the SCDM-cosmology: At early times the angular momenta grow $30\%$ slower in $\Lambda$CDM and $20\%$ slower in classical quintessence with $w_0=-2/3$, with qualitative differences in dark energy models with epoch-varying equation of state $w(a)$.}
\item{Due to the fact that the angular momentum arising in tidal shearing is proportional to the commutator $\matrixx^-=\left[\mathbfss{I},\mathbf{\Psi}\right]$ between the inertia $\mathbfss{I}$ and the tidal shear $\mathbf{\Psi}$ the angular momentum field is sensitive to the relative misalignments of the principal axis systems of the tidal shear and inertia tensors. For that reason both fields are included in the Gaussian random process.}
\item{The angular momentum correlation function and variance is computed from a peak-restricted Gaussian random process, due to the fact that galaxy formation is associated with local peaks in the density field. Technically, the peak restriction introduces a different weighting of the tidal shear-configurations compared to a continuous field. The inertia of a peak region is determined from a local parabolic density profile, and integrated within the boundary given by $\delta=0$, which might be a too coarse approximation in the computation of the inertia, because of the weighting of mass elements with the square of the distance from the centre of gravity.}
\item{The covariance matrices used for describing the relation between the individual Gaussian derivates, are expressed in the $y^n_{\ell m}$-basis (i.e. in spherical coordinates), and transformed to the physical variables with a linear transformation. An important detail is the degeneracy of the density field $\delta$ with the trace of the tidal shear $\sum_\alpha \partial_\alpha\partial_\alpha\Psi=\Delta\Psi$ due to the Poisson equation. For that reason the density field itself, although it is used for deriving the relations between all variables in the Gaussian random process, is not part of the random process itself, but derived from the tidal shear. Consequently the sampling can a priori not be restricted to values $\delta>\nu\sigma_0$, but $\delta$ needs to be recomputed for each sampling point, and eventually be rejected, which results in a sampling efficiency amounting to $\simeq2.5\%$ for the 2-point function.}
\item{Computing the standard deviation from of the distribution of the specific angular momentum $L/M$ confirms the expected scaling $\propto M^{2/3}$ proposed by \citet{1969ApJ...155..393P} and derived by \citet{1988MNRAS.232..339H} in an analytical approach, and provides a valuable check for our code.}
\item{The resulting correlation function $C_{\tilde{L}}(r)$ of the specific angular momentum $\tilde{\bmath{L}}=\vecl/(H_0M_\mathrm{scale})$ (where the rescaling makes it possible to state the angular momentum in units of a single fundamental unit) is determined by the interplay of two fields with different correlation lengths: The inertia $I_{\alpha\beta}$ exhibits short ranged correlations, and the tidal shear $\Psi_{\alpha\beta}$ has the same correlation length as the density field. The resulting correlations in angular momentum have a range of about $1~\mathrm{Mpc}/h$ for Milky Way-sized haloes, and can be fitted well with an empirical formula $C_L(r)\propto \exp(-[r/r_0]^\beta)$.}
\end{enumerate}

Future investigations will include the application of the angular momentum correlation function for computing ellipticity correlation functions, which play an important role in gravitational lensing: There, a common assumption are uncorrelated galaxy shapes, but coupled angular momenta give rise to the effect that the galactic disks of neighbouring galaxies are viewed under similar angles of inclination, such that their ellipticities are naturally correlated. The formalism outlined above can serve to compute the ellipticity correlation function, and furthermore, the intrinsic shape $E$- and $B$-mode correlation functions, $C^E_\epsilon(\ell)$ and $C^B_\epsilon(\ell)$ of the (tensor-valued) ellipticity field $\epsilon$ -- a paper about this topic is in preparation.

\section*{Acknowledgements}
We would like to thank in particular Rob Crittenden, Matthias Bartelmann and Alan Heavens for valuable comments, and Sarah Bridle for organising the workshop on intrinsic alignments. BMS's work is supported by the German Research Foundation (DFG) within the framework of the excellence initiative through the Heidelberg Graduate School of Fundamental Physics, and PMM acknowledges funding from the Graduate Academy Heidelberg as well as from the Graduate School of Fundamental Physics.

\bibliography{bibtex/aamnem,bibtex/references}

\begin{thebibliography}{}

\bibitem[\protect\citeauthoryear{{Abramowitz}, {Stegun} \&
  {Romer}}{{Abramowitz} et~al.}{1988}]{1988AmJPh..56..958A}
{Abramowitz} M.,  {Stegun} I.~A.,    {Romer} R.~H.,  1988, American Journal of
  Physics, 56, 958

\bibitem[\protect\citeauthoryear{{Arag{\'o}n-Calvo}, {van de Weygaert}, {Jones}
  \& {van der Hulst}}{{Arag{\'o}n-Calvo} et~al.}{2007}]{2007ApJ...655L...5A}
{Arag{\'o}n-Calvo} M.~A.,  {van de Weygaert} R.,  {Jones} B.~J.~T.,    {van der
  Hulst} J.~M.,  2007, \apjl, 655, L5

\bibitem[\protect\citeauthoryear{{Bardeen}, {Bond}, {Kaiser} \&
  {Szalay}}{{Bardeen} et~al.}{1986}]{1986ApJ...304...15B}
{Bardeen} J.~M.,  {Bond} J.~R.,  {Kaiser} N.,    {Szalay} A.~S.,  1986, \apj,
  304, 15

\bibitem[\protect\citeauthoryear{{Betancort-Rijo} \&
  {Trujillo}}{{Betancort-Rijo} \& {Trujillo}}{2009}]{2009arXiv0912.1051B}
{Betancort-Rijo} J.~E.,  {Trujillo} I.,  2009, ArXiv e-prints

\bibitem[\protect\citeauthoryear{{Blazek}, {McQuinn} \& {Seljak}}{{Blazek}
  et~al.}{2011}]{2011arXiv1101.4017B}
{Blazek} J.,  {McQuinn} M.,    {Seljak} U.,  2011, ArXiv e-prints

\bibitem[\protect\citeauthoryear{{Catelan}, {Kamionkowski} \&
  {Blandford}}{{Catelan} et~al.}{2001}]{2001MNRAS.320L...7C}
{Catelan} P.,  {Kamionkowski} M.,    {Blandford} R.~D.,  2001, \mnras, 320, L7

\bibitem[\protect\citeauthoryear{{Catelan} \& {Porciani}}{{Catelan} \&
  {Porciani}}{2001}]{2001MNRAS.323..713C}
{Catelan} P.,  {Porciani} C.,  2001, \mnras, 323, 713

\bibitem[\protect\citeauthoryear{{Catelan} \& {Theuns}}{{Catelan} \&
  {Theuns}}{1996a}]{1996MNRAS.282..436C}
{Catelan} P.,  {Theuns} T.,  1996a, \mnras, 282, 436

\bibitem[\protect\citeauthoryear{{Catelan} \& {Theuns}}{{Catelan} \&
  {Theuns}}{1996b}]{1996MNRAS.282..455C}
{Catelan} P.,  {Theuns} T.,  1996b, \mnras, 282, 455

\bibitem[\protect\citeauthoryear{{Catelan} \& {Theuns}}{{Catelan} \&
  {Theuns}}{1997}]{1997MNRAS.292..225C}
{Catelan} P.,  {Theuns} T.,  1997, \mnras, 292, 225

\bibitem[\protect\citeauthoryear{{Chevallier} \& {Polarski}}{{Chevallier} \&
  {Polarski}}{2001}]{2001IJMPD..10..213C}
{Chevallier} M.,  {Polarski} D.,  2001, International Journal of Modern Physics
  D, 10, 213

\bibitem[\protect\citeauthoryear{{Crittenden}, {Natarajan}, {Pen} \&
  {Theuns}}{{Crittenden} et~al.}{2001}]{2001ApJ...559..552C}
{Crittenden} R.~G.,  {Natarajan} P.,  {Pen} U.-L.,    {Theuns} T.,  2001, \apj,
  559, 552

\bibitem[\protect\citeauthoryear{{Croft} \& {Metzler}}{{Croft} \&
  {Metzler}}{2000}]{2000ApJ...545..561C}
{Croft} R.~A.~C.,  {Metzler} C.~A.,  2000, \apj, 545, 561

\bibitem[\protect\citeauthoryear{{Doroshkevich}}{{Doroshkevich}}{1970}]{1970Ap......6..320D}
{Doroshkevich} A.~G.,  1970, Astrophysics, 6, 320

\bibitem[\protect\citeauthoryear{{Efstathiou}, {Bond} \& {White}}{{Efstathiou}
  et~al.}{1992}]{1992MNRAS.258P...1E}
{Efstathiou} G.,  {Bond} J.~R.,    {White} S.~D.~M.,  1992, \mnras, 258, 1P

\bibitem[\protect\citeauthoryear{{Hahn}, {Carollo}, {Porciani} \&
  {Dekel}}{{Hahn} et~al.}{2007}]{2007MNRAS.381...41H}
{Hahn} O.,  {Carollo} C.~M.,  {Porciani} C.,    {Dekel} A.,  2007, \mnras, 381,
  41

\bibitem[\protect\citeauthoryear{{Hahn}, {Teyssier} \& {Carollo}}{{Hahn}
  et~al.}{2010}]{2010MNRAS.405..274H}
{Hahn} O.,  {Teyssier} R.,    {Carollo} C.~M.,  2010, \mnras, 405, 274

\bibitem[\protect\citeauthoryear{{Heavens} \& {Peacock}}{{Heavens} \&
  {Peacock}}{1988}]{1988MNRAS.232..339H}
{Heavens} A.,  {Peacock} J.,  1988, \mnras, 232, 339

\bibitem[\protect\citeauthoryear{{Heavens}, {Refregier} \& {Heymans}}{{Heavens}
  et~al.}{2000}]{2000MNRAS.319..649H}
{Heavens} A.,  {Refregier} A.,    {Heymans} C.,  2000, \mnras, 319, 649

\bibitem[\protect\citeauthoryear{{Heavens} \& {Sheth}}{{Heavens} \&
  {Sheth}}{1999}]{1999MNRAS.310.1062H}
{Heavens} A.~F.,  {Sheth} R.~K.,  1999, \mnras, 310, 1062

\bibitem[\protect\citeauthoryear{{Hirata}, {Mandelbaum}, {Ishak}, {Seljak},
  {Nichol}, {Pimbblet}, {Ross} \& {Wake}}{{Hirata}
  et~al.}{2007}]{2007MNRAS.381.1197H}
{Hirata} C.~M.,  {Mandelbaum} R.,  {Ishak} M.,  {Seljak} U.,  {Nichol} R.,
  {Pimbblet} K.~A.,  {Ross} N.~P.,    {Wake} D.,  2007, \mnras, 381, 1197

\bibitem[\protect\citeauthoryear{{Hirata} \& {Seljak}}{{Hirata} \&
  {Seljak}}{2004}]{2004PhRvD..70f3526H}
{Hirata} C.~M.,  {Seljak} U.,  2004, \prd, 70, 063526

\bibitem[\protect\citeauthoryear{{Hoyle}}{{Hoyle}}{1949}]{1949MNRAS.109..365H}
{Hoyle} F.,  1949, \mnras, 109, 365

\bibitem[\protect\citeauthoryear{{Jing}}{{Jing}}{2002}]{2002MNRAS.335L..89J}
{Jing} Y.~P.,  2002, \mnras, 335, L89

\bibitem[\protect\citeauthoryear{{Jones}, {van de Weygaert} \&
  {Arag{\'o}n-Calvo}}{{Jones} et~al.}{2010}]{2010MNRAS.408..897J}
{Jones} B.~J.~T.,  {van de Weygaert} R.,    {Arag{\'o}n-Calvo} M.~A.,  2010,
  \mnras, 408, 897

\bibitem[\protect\citeauthoryear{{King}}{{King}}{2005}]{2005astro.ph..6441K}
{King} L.,  2005, arXiv:astro-ph/0506441

\bibitem[\protect\citeauthoryear{{Krause} \& {Hirata}}{{Krause} \&
  {Hirata}}{2011}]{2011MNRAS.410.2730K}
{Krause} E.,  {Hirata} C.~M.,  2011, \mnras, 410, 2730

\bibitem[\protect\citeauthoryear{{Lee}}{{Lee}}{2006}]{2006ApJ...644L...5L}
{Lee} J.,  2006, \apjl, 644, L5

\bibitem[\protect\citeauthoryear{{Lee} \& {Park}}{{Lee} \&
  {Park}}{2006}]{2006astro.ph..6477L}
{Lee} J.,  {Park} D.,  2006, astro-ph/0606477

\bibitem[\protect\citeauthoryear{{Lee} \& {Pen}}{{Lee} \&
  {Pen}}{2000}]{2000ApJ...532L...5L}
{Lee} J.,  {Pen} U.,  2000, \apjl, 532, L5

\bibitem[\protect\citeauthoryear{{Linder} \& {Jenkins}}{{Linder} \&
  {Jenkins}}{2003}]{2003MNRAS.346..573L}
{Linder} E.~V.,  {Jenkins} A.,  2003, \mnras, 346, 573

\bibitem[\protect\citeauthoryear{{Mandelbaum}, {Hirata}, {Ishak}, {Seljak} \&
  {Brinkmann}}{{Mandelbaum} et~al.}{2006}]{2006MNRAS.367..611M}
{Mandelbaum} R.,  {Hirata} C.~M.,  {Ishak} M.,  {Seljak} U.,    {Brinkmann} J.,
   2006, \mnras, 367, 611

\bibitem[\protect\citeauthoryear{{Messiah}}{{Messiah}}{1962}]{1962qume.book.....M}
{Messiah} A.,  1962, {Quantum mechanics}.
Amsterdam: North-Holland Publication, 1961-1962

\bibitem[\protect\citeauthoryear{{Peebles}}{{Peebles}}{1969}]{1969ApJ...155..393P}
{Peebles} P.~J.~E.,  1969, \apj, 155, 393

\bibitem[\protect\citeauthoryear{{Regos} \& {Szalay}}{{Regos} \&
  {Szalay}}{1995}]{1995MNRAS.272..447R}
{Regos} E.,  {Szalay} A.~S.,  1995, \mnras, 272, 447

\bibitem[\protect\citeauthoryear{{Sch{\"a}fer}}{{Sch{\"a}fer}}{2009}]{2009IJMPD..18..173S}
{Sch{\"a}fer} B.~M.,  2009, IJMPD, 18, 173

\bibitem[\protect\citeauthoryear{{Schneider} \& {Bridle}}{{Schneider} \&
  {Bridle}}{2010}]{2010MNRAS.402.2127S}
{Schneider} M.~D.,  {Bridle} S.,  2010, \mnras, 402, 2127

\bibitem[\protect\citeauthoryear{{Sciama}}{{Sciama}}{1955}]{1955MNRAS.115....2S}
{Sciama} D.~W.,  1955, \mnras, 115, 2

\bibitem[\protect\citeauthoryear{{Semboloni}, {Heymans}, {van Waerbeke} \&
  {Schneider}}{{Semboloni} et~al.}{2008}]{2008MNRAS.388..991S}
{Semboloni} E.,  {Heymans} C.,  {van Waerbeke} L.,    {Schneider} P.,  2008,
  \mnras, 388, 991

\bibitem[\protect\citeauthoryear{{Sugerman}, {Summers} \&
  {Kamionkowski}}{{Sugerman} et~al.}{2000}]{2000MNRAS.311..762S}
{Sugerman} B.,  {Summers} F.~J.,    {Kamionkowski} M.,  2000, \mnras, 311, 762

\bibitem[\protect\citeauthoryear{{Sugiyama}}{{Sugiyama}}{1995}]{1995ApJS..100..281S}
{Sugiyama} N.,  1995, \apjs, 100, 281

\bibitem[\protect\citeauthoryear{{White}}{{White}}{1984}]{1984ApJ...286...38W}
{White} S.~D.~M.,  1984, \apj, 286, 38

\bibitem[\protect\citeauthoryear{{Zel'dovich}}{{Zel'dovich}}{1970}]{1970A&A.....5...84Z}
{Zel'dovich} Y.~B.,  1970, \aap, 5, 84

\end{thebibliography}
\bibliographystyle{mn2e}

\appendix

\bsp

\label{lastpage}

\end{document}